\documentclass[acs,twocolumn,groupedaddress,reprint,superscriptaddress,showpacs,preprintnumbers,amsmath,amssymb, mathtools]{revtex4}
\usepackage{textcomp}
\usepackage{graphicx}
\usepackage{dcolumn}
\usepackage{bm}%
\usepackage[colorlinks=true,citecolor=blue]{hyperref}
\hypersetup{colorlinks=true,citecolor=blue,linkcolor=red,urlcolor=blue}
\newcommand {\be}{\begin{equation}}
\newcommand {\ee}{\end{equation}}

\usepackage{ulem}
\usepackage{xcolor}

\begin{document}
\title{An ab initio supercell approach for high-harmonic generation in liquids}

\author{Zahra Nourbakhsh}
\email[]{zahra.nourbakhsh@mpsd.mpg.de}
\affiliation{Max Planck Institute for the Structure and Dynamics of Matter, Luruper Chaussee 149, 22761 Hamburg, Germany.}

\author{Ofer Neufeld}
\affiliation{Max Planck Institute for the Structure and Dynamics of Matter, Luruper Chaussee 149, 22761 Hamburg, Germany.}

\author{Nicolas Tancogne-Dejean}
\affiliation{Max Planck Institute for the Structure and Dynamics of Matter, Luruper Chaussee 149, 22761 Hamburg, Germany.}

\author{Angel Rubio}
\email[]{angel.rubio@mpsd.mpg.de}
\affiliation{Max Planck Institute for the Structure and Dynamics of Matter, Luruper Chaussee 149, 22761 Hamburg, Germany.}
\affiliation{The Hamburg Centre for Ultrafast Imaging, Hamburg, Germany.}
\affiliation{Nano-Bio Spectroscopy Group and ETSF, Departamento de Fisica de Materiales, Universidad del Pa\'is Vasco UPV/EHU, 20018, San Sebasti\'an, Spain.}
\affiliation{Center for Computational Quantum Physics (CCQ), The Flatiron Institute, 162 Fifth Avenue, New York, New York 10010, USA.}

\begin{abstract}
Many important ultrafast phenomena take place in the liquid phase. 
However, there is no practical theory to predict how liquids respond to intense light. 
Here, we propose an $ab~initio$ accurate method to study the non-perturbative interaction of intense pulses with a liquid target to investigate its high-harmonic emission. 
We consider the case of liquid water, but the method can be applied to any other liquid or amorphous system. 
The liquid water structure is reproduced using Car-Parrinello molecular dynamics simulations in a periodic supercell.  Then, we employ real-time  time-dependent density functional theory to evaluate the light-liquid interaction.
We outline the practical numerical conditions to obtain a converged response.
Also, we discuss the  impact of nuclei ultrafast dynamics on the non-linear response of system. 
In addition, by considering two different ordered structures of ice, we show how harmonic emission responds to the loss of long-range order in liquid water.
\end{abstract}

\maketitle

\section{Introduction}\label{intro}
High-harmonic generation (HHG) occurs as a result of the interaction between an intense infrared laser pulse with a target matter where light at integer multiples of the incident pulse frequency is emitted \cite{hhg}.  
HHG is a source of coherent table-top extreme-ultraviolet (XUV) radiation with direct applications for attosecond technology and exploring 
electron ultrafast dynamics in materials \cite{as,nour}. 
 HHG was discovered in 1977 \cite{ghhg1}; it has been extensively studied in atomic and gaseous systems 
as well as crystals and condensed matters \cite{hhgsol}. Due to the higher density of solids and their lattice periodicity, solids are potential materials to emit intense high harmonics; for instance, monoatomic crystals show a brighter HHG spectrum in comparison with their gas phases under the same incident pulse \cite{vdw-solids}. However, the pulse intensity in solids is limited to the  target damage threshold. Since liquids are condensed systems that, like gases, can tolerate much higher intensities, they could be advantageous as novel sources of attosecond pulses and XUV. Moreover, liquids are attractive systems as they are the platform of most chemical and bio-chemical reactions \cite{pr1}.

In gases, HHG arises from an ensemble of independent isolated emitters, and the HHG process is well explained by the semi-classical three-step model \cite{hhgas1,hhgas2}. On the contrary, in periodic solids HHG reflects the crystalline symmetries and is well described by the contribution of interband and intraband dynamics \cite{hhgsol, sc1,sc2,sc3}, which are based on the lattice periodicity.
Beside these systems, there are several experimental and theoretical studies on HHG from the non-periodic systems like amorphous \cite{amor1, amor2} and defected solids \cite{defect1, defect2, defect3,defect4,defect5}. We therefore expect, based on these different works, that liquids are behaving differently from gases and solids, and it is one of the goal of the present work to understand how liquids are different from other condensed-matter phases.

Following more than a decade of progress \cite{wd1, wd2,wd3,wd4,wd5,wd6}, HHG measurements on the bulk of water and several alcohols revealed the different response of liquids to intense pulses \cite{wl}. This experiment displayed the strong effect of molecular electronic structure on HHG;
 moreover, using one-dimensional semiconductor Bloch equations, this work shows the impact of system bandgap and bandwidth on HHG.
Another work, based on a one-dimensional  model, attempted to clarify some ambiguities in the liquid response to strong laser fields \citep{prll}. 
However, HHG in liquids remains a challenging field of research which is not well-understood, and several crucial questions remains to be answered before a clear picture could be drawn about dynamics at place when intense lasers irradiate liquids and the resulting harmonic emission.
In order to bridge this gap, we will here introduce an approach to evaluate nonlinear light-liquid interactions on the basis of $ab~initio$ methods  and we use it to address some of these important questions.
In addition to this work, in another study \citep{cluster} we proposed a faster computational method based on a cluster approach to compare the HHG response of different polar and nonpolar liquids, to have a better understanding of electronic structure of liquid on HHG response.
The cluster approach has no periodicity and therefore requires only one K-point, which speeds up the calculations by 2-3 orders of magnitude; however, surface contributions must be suppressed in this approach, and it is also much more difficult to study molecular and ionic dynamics with it.
 Despite the inclusion of some simplifications, the cluster method also provides good predictions for the HHG response of liquid systems; Supplementary Fig.~S7 shows a comparison between the results of these two methods.

\begin{figure}
	\centering
	\includegraphics[width=0.6\linewidth]{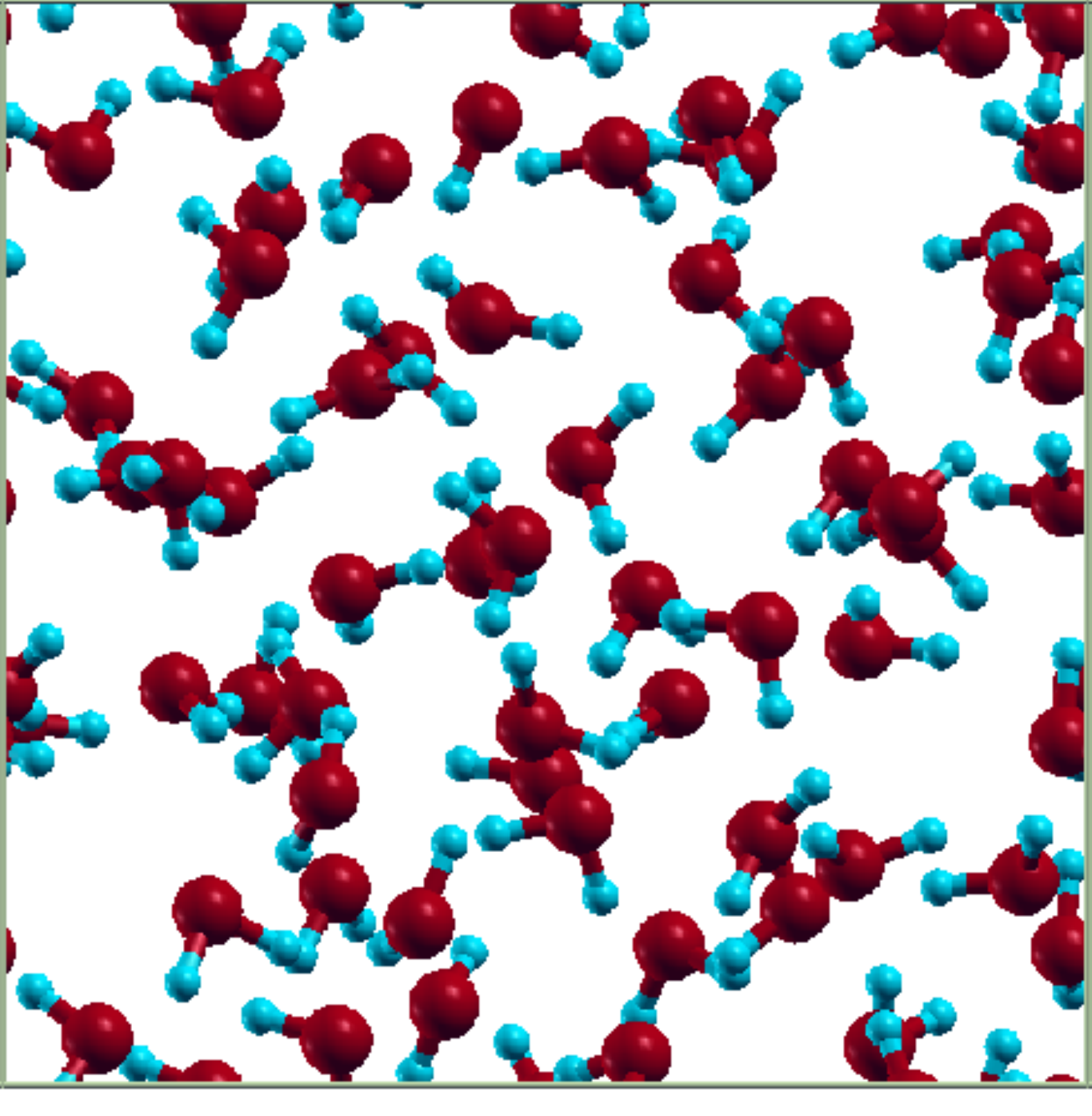}
	\caption{\label{cell} {\bf A typical considered structure of liquid water.} It contains 64 water molecules in a cubic cell with a length of 12.43 \AA.}
\end{figure}

\section{Results}\label{result}

In this work, we employ real-time  time-dependent density functional theory (TDDFT) \cite{tddft1, tddft2} 
to calculate harmonic emission from liquid water as a benchmark example; however, the proposed method is applicable to any other liquid and amorphous system. In this regard, we first simulate the structural and dynamical behavior of water \cite{wdft, scan, wdft2, wdft3, wdft4} 
with Car-Parrinello molecular dynamics (CPMD)~\cite{cpmd,cp2} in a periodic cubic supercell, as illustrated in Fig.~\ref{cell}, including 64~H$_2$O molecules using the experimental liquid water density, see Supplementary Figure S1. 
After successfully reproducing the liquid water configuration, the non-linear interaction of strong laser fields with liquid water is determined using microscopic TDDFT for several different configurations of liquid water supercell, which are obtained from the CPMD to mimic the experimental situation. The isotropic response of the full system is recovered
when a sufficient number of configurations of liquid are used, as explained below. 
Note that the coupling of CPMD and real-time TDDFT operate on different timescales, where TDDFT describes the response of the system on femtosecond timescales, whereas CPMD manages to capture the thermodynamics features of the liquid on picosecond timescales. 
More details on the methodology, as well as on the simulation parameters, are given in Sec.~\ref{method}.

\subsection{HHG from liquid water} \label{whhg}

\begin{figure}
	\centering
	\includegraphics[width=1.\linewidth]{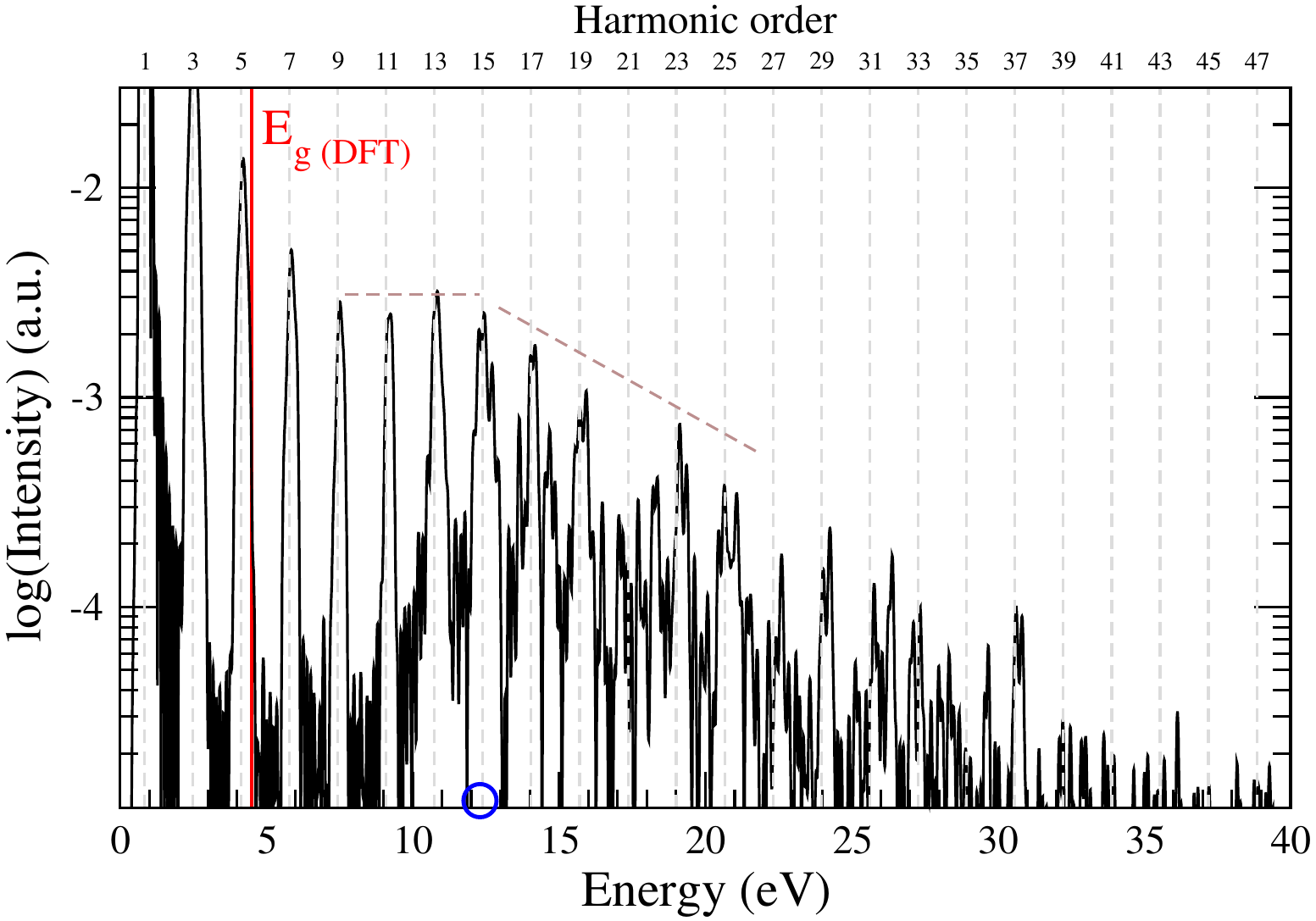}
	\caption{\label{lhhg} {\bf The calculated HHG spectrum along with the incident pulse ($x$-axis) from liquid water.}  The gray dashed lines show the odd harmonic frequencies. The blue circle marks the cutoff energy. The solid red line indicates the liquid water DFT energy gap (4.6~eV) averaged over several subsystems. The brown dashed lines mark the plateau and the decay area.}
\end{figure}

\begin{figure*}
	\centering
	\includegraphics[width=1.\linewidth] {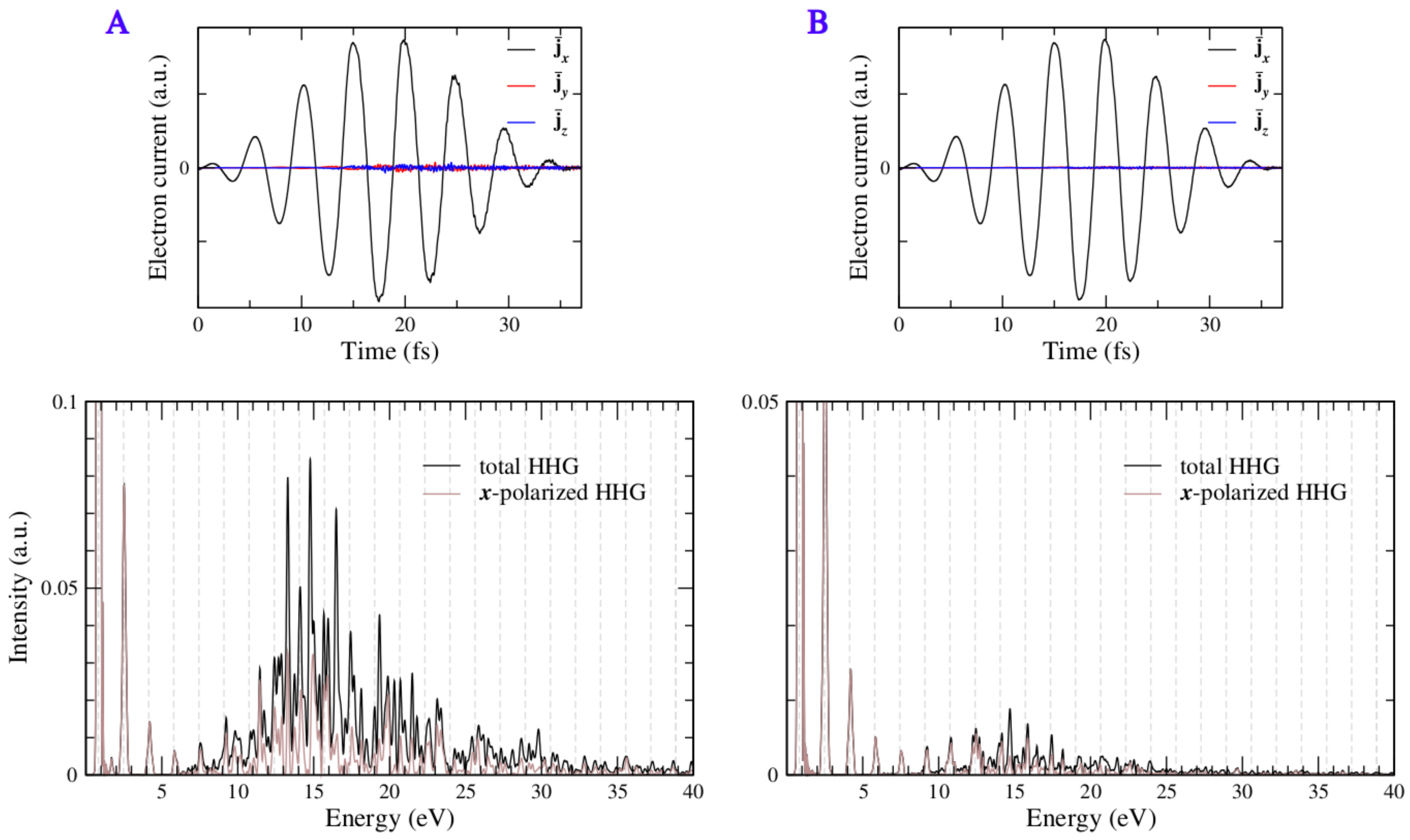}
	\caption{\label{nimp} {\bf Toward the isotropic limit: Impact of increasing the number of subsystems on arising the isotropic properties of liquid water.} The calculated time dependent electron current per water molecule in the liquid water in the direction along with the incident pulse ($x$-direction) as well as perpendicular directions averaged over (A) 64 molecules of one subsystem, and (B) 640 molecules of ten subsystems. 
	The bottom panel shows the corresponding total HHG spectrum (HHG$_{tot}$) as well as HHG with polarization aligned with the pulse polarization (HHG$_{x}$); HHG$_{tot}$ = HHG$_{x}$+HHG$_{y}$+HHG$_{z}$. The gray dashed lines indicate the odd harmonic frequencies.}
\end{figure*}

Using the above-described method, we computed the high-harmonic response of liquid water to a typical laser pulse used in experiment~\cite{wl} (a wavelength of 1.5 $\mu$m, corresponding to a photon energy of 0.83~eV, and electric-field strength in the bulk of approximately 1.2~V/\AA~,corresponding to the intensity of a few 10 TW/cm$^2$). The pulse duration at full width at half-maximum (FWHM) is equal to 18~fs with a sine-squared envelope shape for the vector potential; the carrier-envelope phase of the pulse vector potential is set to zero.
Figure~\ref{lhhg} presents our calculated HHG spectrum for liquid water interacting with the incident laser pulse. 
This spectrum is obtained by averring over many CPMD configurations, as explained below.

The HHG spectrum, shown in Fig.~\ref{lhhg}, exhibits odd harmonics only, up to order fifteen. 
 Importantly, we obtain that the spectrum exhibits three clear regions. The first region shows an exponential decay of the harmonic yield up to harmonic order 9, and corresponds to what is usually called perturbative harmonics in HHG from gases. Then, unlike most solids, we obtain a clear plateau with the cutoff energy of 12.4~eV.  At the end of the plateau, an exponential decay of the yield is again observed. We note that above the harmonic order 15 the presence of odd and even harmonics. This unphysical even harmonics are a consequence of the finite number of subsystems used in the averaging procedure, as explained in Sec.~\ref{isotropic}.\\
We note that the spectrum displayed in Fig.~\ref{lhhg} results from averaging the HHG response over 85 different subsystems (5440 water molecules), and therefore represent a formidable numerical challenge even for today computing resources.
As displayed in Supplementary Fig.~S2 in  the  Supplementary  Materials, 
our spectrum is appropriately converged for the energies below the energy cutoff. \\

Compared to the experimental measurement of HHG from liquid water reported in Ref.~\cite{wl}, our simulated spectrum nicely reproduces different features, such as absence of even harmonics, the cutoff energy, and the appearance of the plateau. We  also point out that we have recently employed our numerical approach to understand the impact of incident pulse characteristics on HHG from liquid water, and obtained that in agreement with experiment, liquid HHG cutoff energy is independent from the driving wavelength and it is weakly sensitive to the pulse amplitude \cite{landa}.  
These behaviors are different from the atomic and molecular gases where the ponderomotive energy $U_p \propto \lambda^2 E^2$ defines the scaling of the cutoff energy  with respect to the pulse amplitude and wavelength. Also, it differs from solid systems where the cutoff can scale linearly with the field amplitude \cite{nour,hhgsol}. However, the independence of cutoff energy from pulse wavelength has been reported for different solids \cite{hhgsolT1,explanda}, too.
 We also note that the presence of a clear plateau is usually not observed in bulk solids, with the notable exception of rare-gas solids, which even exhibit multiple plateaus\cite{vdw-solids}.\\
Overall, these observation points toward the fact that liquids behave differently than both bulk solids and atomic and molecular gases, even if they exhibit some specificity that can resemble what is found in either the former or the latter one.

\subsection{Toward the isotropic limit}\label{isotropic}
Because of the asymmetric shape of water molecules, both odd and even harmonics can be generated along and perpendicular to the incident pulse polarization direction when one performs a simulation for a finite number of molecules. However, since the full bulk liquid is isotropic, all the emission perpendicular to the incident pulse direction as well as even harmonics must cancel out \cite{evh1,evh2}.  With this respect, the liquids are behaving  as amorphous solids, which do not exhibit even harmonics~\cite{amor1}.
This fundamental property needs to be restored in the simulations, and is indeed obtained by our calculations as the response builds up over a sufficiently large number of liquid configurations (subsystems).
Figure~\ref{nimp} shows the time evolution of a typical electronic current per H$_2$O molecule of liquid water and the corresponding HHG spectra driven by an intense linearly polarized along the $x$ direction, using the same laser parameters as given above. This figure compares the results averaged over one subsystem (displayed in \ref{nimp}\textbf{A}) and ten subsystems (displayed in \ref{nimp}\textbf{B}).
According to Fig.~\ref{nimp}\textbf{A}, the spectrum in the low energy region (corresponding to perturbative harmonics) cleanly presents odd-only harmonics with polarization along with the incident pulse, as expected for the response of an isotropic material. However, as shown by the electronic current, the two other components exhibit fast oscillations, indicating non-vanishing contributions from these directions from high-order harmonics, whereas these contributions are expected to be zero in the isotropic limit. This gives us a simple measure of the convergence of the HHG spectrum with respect to the number of subsystems. When the total spectrum (summed over all Cartesian directions) becomes the same as the one along the laser direction, the spectrum is considered to be converged. Clearly, above 6\,eV, the HHG spectrum for a single subsystem is not isotropic anymore. To converge the higher harmonics more subsystems are needed.
Figure~\ref{nimp}\textbf{B}, obtained by averaging over 10 subsystems, clearly shows that the expected isotropic response is correctly
improved with the addition of more subsystems, as all transverse components of the current  are strongly reduced. Moreover, the even harmonics, present in Fig.~\ref{nimp}\textbf{A} starting from ~8.5eV are now only present above 13\,eV.
Importantly, we note that adding more subsystems leads to a decrease of the harmonic yield. This indicates the absence of coherence in between the different subsystems, and is similar to the effect observed when comparing crystalline and amorphous solids~\cite{amor1}.\\
The isotropicity of the HHG spectra in a particular energy region is good indication that the results are numerically converged.
The required number of subsystems for convergence depends on the target system and its molecular symmetries \citep{cluster}  as well as pulse characteristics such as  pulse wavelength and amplitude.
Nonetheless, we find that in order to converge the HHG spectrum up to 15~eV, and hence above the energy cutoff of liquid water for our driving conditions,  we need approximately 85 subsystems, corresponding to 5440 water molecules. The corresponding spectrum and the comparison between the total HHG spectrum and the one along the $x$ direction only is shown in Supplementary Fig. S2.\\

\begin{figure*}[t]
	\centering
	\includegraphics[width=1.\linewidth]{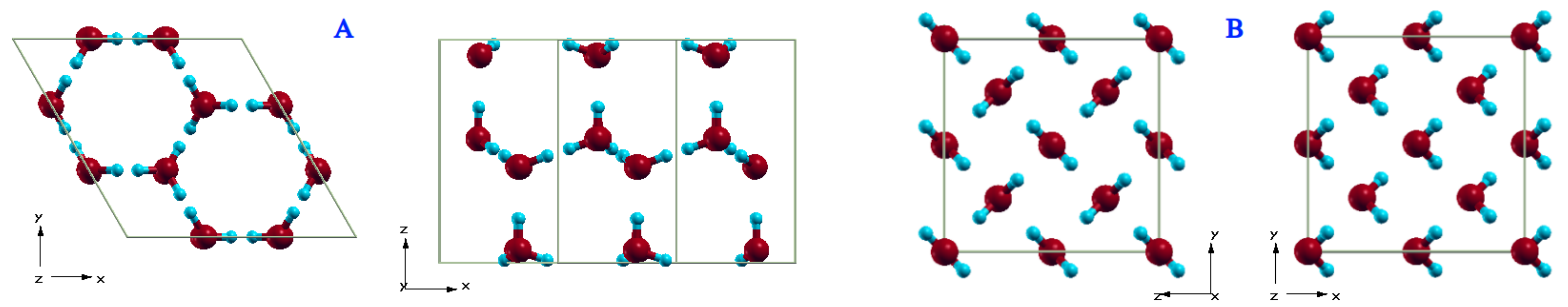} 
	\caption{\label{struct} {\bf Top view and side view of crystalline ice in (A) hexagonal (Ih) and (B) cubic (Ic) structures.} The distance between neighboring oxygen atoms along the hydrogen bond is approximately 2.7~\AA. The hexagonal structure has 12 basis molecules in its unit cell while the cubic lattice has just 2 basis molecules with fcc symmetry. 
		In the cubic structure, the oxygen atoms are located on the same positions as C atoms in diamond but the O-H bond directions are toward the neighboring molecule O atoms to fulfill the hydrogen bonding.}
	\centering
	\includegraphics[width=1.\linewidth]{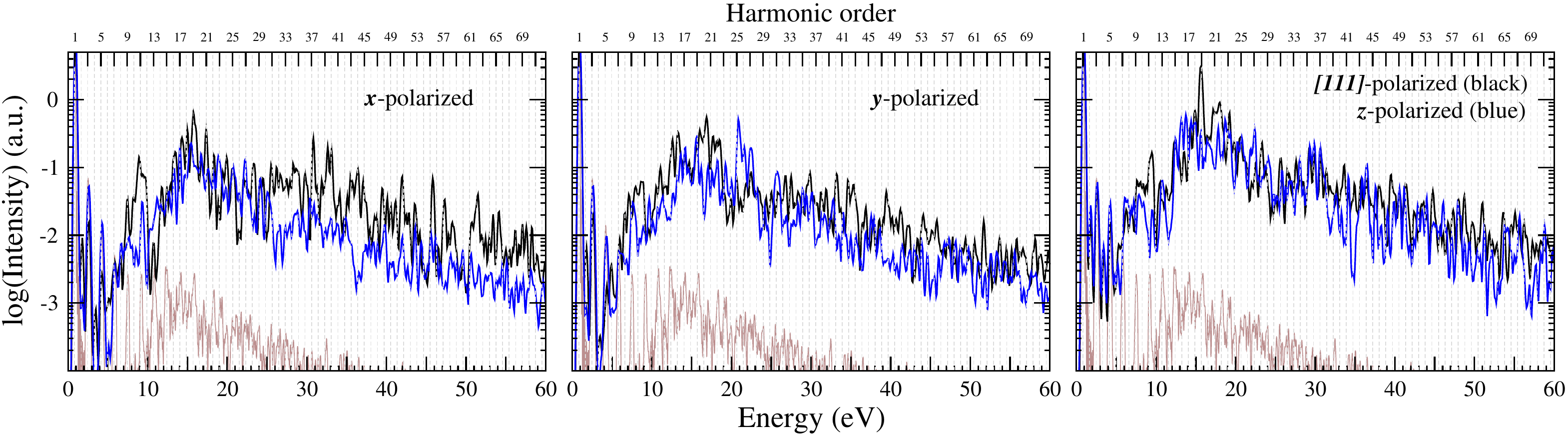} 
	\caption{\label{aniso} {\bf Impact of periodicity in harmonic emission from water.} HHG from cubic (hexagonal) structure is shown in black (blue), and the liquid water spectrum is displayed in light brown. We use a Gaussian filtering to smooth the ice spectra, with a width taking as $\omega/6$, where $\omega$ is incident pulse frequency.} The differences between the curves is discussed in the main text.  
\end{figure*}

\subsection{Environmental impacts on HHG: Ordering, periodicity, and anisotropy}\label{enviro}

We now show how HHG spectral features respond to the structural arrangement and breaking the long-range order in water  by comparing to HHG from ordered phases of ice. 
We consider two well-known ordered phases of ice, namely Ih and Ic, shown in Fig.~\ref{struct}. In order to compare their emission with the liquid phase, we slightly compress these structures (approximately, 2-3\% in each direction) to be in the same density as liquid water. Supplementary Figure~S4  shows the electronic density of states of these crystalline structures as well as liquid phase; this figure predicts the DFT energy gap of approximately 5.9~eV for both crystals which is $\sim 1$~eV higher than the DFT energy gap of liquid water (at room temperature).  

Figure~\ref{aniso} shows the harmonic emission from crystalline Ic and Ih structures and compares them with liquid water HHG. 
All the spectra in Fig.~\ref{aniso} arise from the same driving pulses, but to see the impacts of crystal anisotropy and molecular alignment on HHG, we consider three different pulse polarizations along the high-symmetry directions of each crystal. 
Since inversion symmetry is broken for these crystalline structures, even harmonics also appear in their spectra.
As seen in Fig.~\ref{aniso}, for the low-harmonics up to order 7, emission yield from liquid water is comparable or even higher than those of crystalline phases. This is expected for perturbative harmonics, as in our simulations the energy gap of bulk ice is higher than for liquid water, and hence the excitation of carriers is more important in water than ice.\\

But, for the higher harmonics, because of disturbing the emission coherency in the disordered liquid phase, the liquid phase HHG  strongly decreases. This behaviour is similar to the experimental measurements of the amorphous phase of quartz, namely fused silica, which shows significantly weaker harmonic emission and lower cutoff energy in comparison to crystalline quartz \cite{amor1, amor2}. 
However, not only the harmonic yield is stronger in the two phases of ice, but in sharp contrast with the liquid phase and usual solids, the harmonic emission in the solid ice raises strongly above the bandgap. We attribute this effect to the dominate role of the interband contribution, which can only occur above the bandgap. As the band curvature in ice phases is quite small, because of the ionic nature of the bounds, the intraband contribution is expected to be smaller than the interband one, which would explain the sharp rise above the band edge, which is not usually seen in covalent-bound semiconductors.
\\

In liquids, the extension of the plateau (the energy cutoff) is limited by the scattering and the mean-free path of the low-energy electrons in the liquid~\cite{landa}.
We find here that a strong harmonic emission is observed in both considered phases of ice.
The cutoff energies in all ordered phases are higher than the cutoff of liquid  and extend up to approximately 20 eV. This reflects the fact that in solids, electrons can coherently move up the Brillouin zone edge, where electrons can scatter to higher conduction bands~\cite{wang2020role}. This coherent motion of the electrons in the solid phase is evidenced by a strong harmonic emission around 15\,eV. Indeed, both Ic and Ih density of states (displayed in Fig. S4) show Van Hove singularities at approximately 15~eV.
The van Hove singularities are known to lead to a strong enhancement of the harmonic yield of the interband current~\cite{uzan2020attosecond}. We therefore link the strong harmonic emission around 15\,eV to the presence of van Hove singularities at this energy in the joint density of state of Ic and Ih ice. This is also another evidence of the dominant role of the interband mechanism in the harmonic emission in ice.\\
This feature is observed for all the three polarization considered here, even if strong emission at different energies are also observed (e.g. around 9\,eV or around 21\,eV ) for some polarizations or phases. These differences are dictated by the details of the bandstructure of Ih and Ic ices, which we are not focusing on in the present work.\\
These results reveal how different the mechanisms responsible for HHG in liquid and solids can be. In bulk ice, we observe a strong influence of the interband current when electrons are coherently steered by the laser from the Brillouin zone center to its edge, and of band-structure effects (van Hove singularities). Meanwhile, in liquids, the harmonic response mostly reveal the importance of the mean-free path on the electron trajectories in real-space, which limit the excursion of the electron and hence the extension of the plateau.

In  Fig.~\ref{aniso}, because of the higher number of basis molecules in Ih crystal compared to the Ic crystal, the emission from the disordered structure of liquid water in the low energy region is more similar to the Ih emission; note that the low  harmonics arise mostly from the single-molecule perturbative response.
For the higher harmonics, due to the fewer basis molecules of Ic structure and therefore its higher ordering and coherency, its harmonic emission is stronger than Ih structure.
Notably, for both crystalline phases, the HHG is enhanced when the light polarization direction is along the hydrogen bonds connecting two neighboring molecules.
In this direction, the gradient of the ionic potential is strongest since it aligns with the O$^{2-}$-H$^+$...O$^{2-}$ ions; and it causes  significant enhancement of harmonic emission \cite{hhgsolT1}.

We end this part by discussing the impact of electron-electron interactions on HHG. Due to weaker screening in the low dimensional systems, like two-dimensional materials or isolated molecules,
electron-electron interactions in these systems strongly affect the harmonic emission (for example see Ref.~\cite{hbn}).
To evaluate the role of the e-e interactions in liquid or crystalline water, we calculate harmonic spectra based on the independent particle  approximation;
our results, presented in Fig.~S5 in  the  Supplementary  Materials,  
reveal that independent-particle approximation gives an acceptable view from the overall features of HHG spectral in the bulk systems. 
It thus confirms the qualitative validity of the other studies on HHG from liquid water with simpler models based on independent particles \cite{cluster, landa}.

\subsection{Impact of ion dynamics on HHG}\label{ion-motion}

\begin{figure}
	\centering
	\includegraphics[width=1\linewidth]{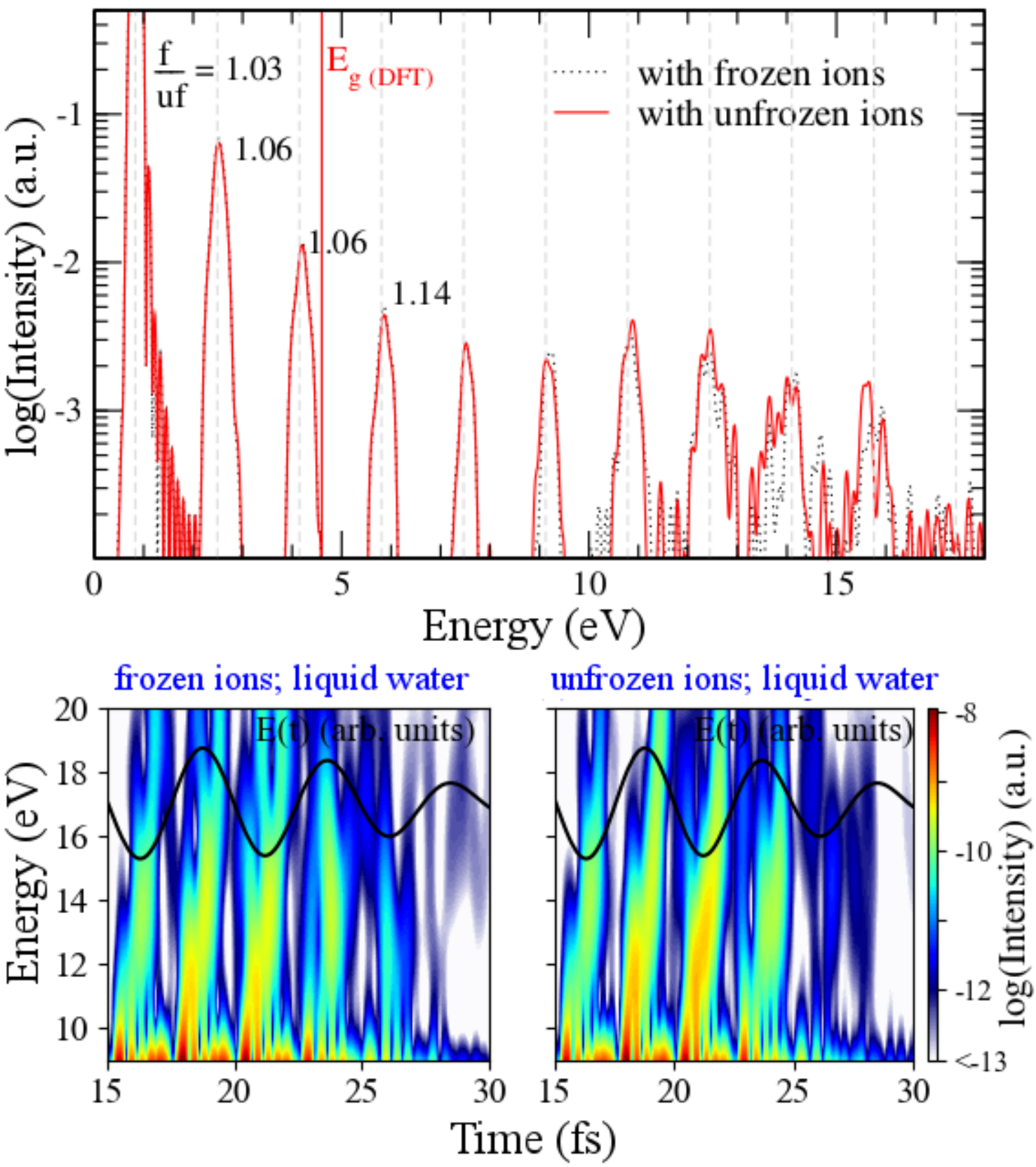}
	\caption{\label{ionmov} {\bf Impact of the ionic motion on HHG from liquid water.} The initial velocities are taken from CPMD. The related time-frequency analysis are shown in the bottom panel; this figure also shows the time profile of the electric field.	}
\end{figure}

As we show in the previous section, unlike gaseous systems, the relative positions of atoms or molecules are important in the HHG response from condensed systems like liquid water. This can be understood by the fact that once an electron is excited in liquid water, the resulting hole is delocalized over the neighboring molecules \cite{hj}. This effect is even more pronounced in order phases like ice. Since the dynamics of the center of the molecules as well as the proton transfer under strong laser field are fundamental characteristics of liquids, and in particular of water,
a relevant question is whether or not the HHG features are modified by the ionic motion during the interaction of our ultrashort intense pulse with the system, and if yes, up to which extend. We note that with our previous cluster approach \citep{cluster} it is much less natural to investigate these effects. \\
Figure~\ref{ionmov} shows the impact of ultrafast ion dynamics on harmonic emission from liquid water; we also discuss it for Ih crystal in the Supplementary.  
This figure compares the HHG response calculated in the same conditions, with and without including the dynamics of the nuclei, described at the level of Ehrenfest dynamics.
At first glance, our results imply that ion dynamics do not change the overall features of HHG spectrum.
For the first four  peaks, 
Fig.~\ref{ionmov} reports the ratio of the harmonic yield when the ions are frozen to that of allowed ionic motion. As shown this ratio is greater than one and increases with harmonic energies which is in agreement with the similar finding in the gas phase for different isotopes of water where the heavier D$_2$O isotope with slower ion dynamics and slower vibration compared to H$_2$O shows stronger HHG increasing with harmonic frequency \cite{vibhhg}.  
In addition, other experimental \cite{nucdyn} and theoretical \cite{lein} studies comparing HHG from the D$_2$ and H$_2$ isotopes reported a similar behavior. \\

The fact that only the lowest harmonics are affected could be understood if one consider the averaged electronic structure of liquid water in terms of a bandstructure.  The  ion dynamics on the hydrogen bonds mostly affect the bottom of the first conduction band (see Supplementary Fig. S3).  This implies that the intraband motion, leading to below bandgap harmonics, and interband current from the first conduction band are modified, modifying the first harmonic orders. This trend does not continue for the harmonics above the gap and  the plateau region, in agreement with the fact that higher conduction bands have a dominant oxygen orbital character. This explains why the plateau region, roughly up to harmonic order 13, is less sensitive to the dynamics. Around the energy cutoff, the effect of the ionic motion is more pronounced. This part of the spectrum is ultimately dictated by the mean-free path of the electrons in the liquid~\cite{landa}, the dynamics of the molecules themselves starts to play an important role, explaining why differences are again observed. The time-frequency analysis of these spectra, discussed below gives further hints on how the effect of the nuclear dynamics builds up in time.\\

 Let us comment here on the use of the Ehrenfest dynamics in our simulations. The Ehrenfest dynamics has been shown to be capable of describing the dissociation of water molecules under electric fields~\cite{PhysRevLett.108.207801} and is therefore capable, at least partially, to capture the proton transfer effects, which are an highly important aspect of the ion dynamics in water.  
However, it is important to note here that the Ehrenfest dynamics employed in our calculations ignores nuclear quantum mechanical, which are known to be important in gases, as they lead for instance to decoherence effects. 
In fact the quantum nuclear effects have also been shown to be important in liquid water~\cite{cassone2020nuclear}. However, because of the delocalization of the hole~\cite{hj}, we expect that effect stemming from the classical motion of the ions should dominate over intra-atomic and intra-molecular nuclear quantum-mechanical effects for what concern the HHG spectra, in particular for the overlap of the ground-state wavefunction and the excited wave-packet. Indeed, we are in a regime of laser intensity for which the force exerted by the laser on the ions is quite important. Our results already point towards the dominant role of the hydrogen bond motion and allow us to quantify how much the classical motion of the ions affects the spectra. This is a first hint towards understanding the complete effects of ionic motion on the ultrafast electron dynamics in liquids driven by intense laser fields.

Our time-frequency analysis, shown in the bottom panels of Fig.~\ref{ionmov}, indicates that despite the similar trends of HHG spectra, 
the attosecond characteristics of liquid water, such as the chirp of the electron wavepackets, is affected by the ultrafast dynamics and vibrations of the molecules.  Most of the observed differences in the area found around the cutoff energy, in the plateau region. The motion of the ions is for instance found to suppress the emission events above 25\,fs. In fact, while harmonic emission before 20\,fs is found to be mostly not affected by the ion dynamics, later times display a modified dynamics.
A similar picture can be obtained for ice (see Supplementary Fig. S6) and again, most differences are found to occur after 20\,fs of dynamics.
This reveals that while short laser pulses will not be sensitive to the ionic motion, which is not significant enough during the first few femtosecond to lead to sizable changes, a longer laser pulse will critically depend on the dynamics of the molecules, and a theoretical modelling needs to take this aspect into account. Our simulation therefore indicate the range of validity of the frozen-ion approximation, at least for our laser parameters.

\section{Discussion}

In summary, this work presents an accurate method to describe the interactions between light and liquids on the basis of coupling CPMD and TDDFT simulations.  
We applied this method for the case of liquid water to study its nonlinear response to intense ultrashort infrared pulses; but, it can also be employed for other liquids or amorphous systems. 
Our method  predicts different experimental features of liquid water HHG, such as absence of even harmonics, cutoff energy and plateau behavior. 
We show that liquid water generate odd harmonics up to the 15th order in the range of extreme ultraviolet wavelengths.
The weak impact of electron-electron interactions on liquid water HHG is marked.
By considering two ordered structures of ice crystal, we show how lacking the long-range order in liquids, same as amorphous solids  \cite{amor1, amor2}, reduces its harmonic emission, and the comparison between bulk ice phases and liquid water reveals the difference in the microscopic mechanism  for the harmonic emission. 
Our calculations also reveal the impact of the ultrafast ion dynamics of liquid water and how it affects the attosecond response of system.
These results deepen our understanding or electron dynamics in liquids, revealing how HHG from liquids is different from order phases of ice.

\section{Materials and Methods}\label{method}

\subsection{Molecular dynamics simulations} \label{cpmet}

The CPMD \cite{cpmd, cp2} is an extension of the Lagrangian formalism of classical molecular dynamics, which uses density functional theory (DFT) total energy to describe the interactions among atoms. This Lagrangian is as bellow (the equations are written in atomic units)
\be
\label{lag}
\begin{split}
\mathcal{L} = \frac{1}{2} \sum_I^{\mathrm{}}\ M_I\dot{\mathbf{R}}_I^2 + \mu\sum_i^{\mathrm{}}\int d\mathbf r\ |\dot{\phi}_i(\mathbf r,t)|^2  \\
- E\left[\{\phi_i\},\{\mathbf R_I\}\right]+\sum_{ij} \lambda_{ij} \left( \int d\mathbf r \phi^{*}_i(\mathbf r,t) \phi_j(\mathbf r,t) -\delta_{ij} \right).
\end{split}
\ee
The first term in Eq.~\ref{lag} shows the kinetic energy of nuclei; $M_I$ and $R_I$ denote the nuclei masses and their positions, respectively. The second term couples electronic degrees of freedom to the ionic motion by defining a fictitious kinetic energy, where $\mu$ is a fictitious mass for electrons and $\phi$ is the Kohn-Sham electronic wavefunction; 
this term keeps the electrons close to the ground state of the updated configurations during the dynamics, and avoids the costly self-consistent total energy minimization at each time step.
The third term in Eq.~\ref{lag} is the Kohn-Sham DFT total energy. Finally, the matrix $\lambda_{ij}$ in the last term defines a set of Lagrange multipliers to ensure orthonormality of the wavefunctions.
The Euler-Lagrange equations of motion derived from Eq.~\ref{lag} are as follows:
\be
\label{eom}
\begin{split}
M_I \ddot{\mathbf R}_I = - \nabla_I \, E\left[\{\phi_i\},\{\mathbf R_I\}\right],
\\
\mu \ddot{\phi}_i(\mathbf r,t) = - \frac{\delta E}{\delta \phi_i^*(\mathbf r,t)} + \sum_j \lambda_{ij} \phi_j(\mathbf r,t).
\end{split}
\ee

In this article, the CPMD calculations to produce liquid water structure are carried out with the Quantum-Espresso package \cite{qe} in the canonical ensemble at the temperature of 300~K.
In our molecular dynamics (MD) simulations, we use the fictitious mass of 100~a.u. for the electrons, a time step of 2~a.u. and frequency of 60~THz for the Nose-Hoover thermostat \cite{nosehover,cp3}.
The Nose-Hoover thermostat provides the energy exchange with the heat bath by considering an additional degree of freedom and modified the ionic equation of motion defined in Eq.~\ref{eom} as $M_I \ddot{\mathbf R}_I = - \nabla_I \, E\left[\{\phi_i\},\{\mathbf R_I\}\right] - M_I \dot{\mathbf R}_I \dot\eta$ where
$Q \ddot{\eta} = \big( \sum_I^{\mathrm{nuclei}}\ M_I\dot{\mathbf{R}}_I^2 -g k_B T  \big) $, Q is thermostat mass (considered 60~THz), $g$ is number of ionic degrees of freedom, $k_B$ is Boltzmann constant and T~=~300~K is the temperature of the thermostat \cite{cp3}.

Note that Van~der~Waals (vdW) interactions play an important role
in the water structure \cite{wl, wdft, scan,  wdft2, wdft3, wdft4}.
Since the local density and generalized-gradient approximations (LDA and GGA), do not properly include the vdW interactions \cite{wdft}, they overestimate the hydrogen bondings and fail to reproduce the structural properties of liquid water \cite{wdft, scan, wdft2, wdft3, wdft4}.
Here, we use norm-conserving GGA-revPBE \cite{revpbe} pseudopotentials for oxygen and hydrogen atoms and Grimme-D2 dispersion correction to describe the non-local correlation effects.

Figure~S1 in  the  Supplementary  Materials shows our radial distribution function between oxygen atoms computed  from our periodic cubic system with the unit cell of 12.43~\AA~ including 64  H$_2$O molecules  over 20~ps CPMD simulations. As this figure shows, our simulation
is in good agreement with the experimental neutron scattering \cite{neutron-exp} and x-ray diffraction \cite{xray-exp} measurements. So our liquid water structure is simulated properly.

\subsection{Ultrafast dynamics simulations}\label{tdft}

TDDFT calculations of the time evolution of electronic wave functions under the influence of an external laser field are performed with the Octopus code \cite{oct} on the basis of the  Kohn-Sham equation, defined as
\be
\begin{split}
\label{kseq}
	i \dot{\phi}_i(\mathbf{r},t) =
	\Big(-\frac{\nabla^2}{2} \mathbf + v_{ext}(\mathbf{r},t) + v_H[n(\mathbf{r},t)] \\
	+ v_{xc}[n(\mathbf{r},t)] \Big)  \phi_i(\mathbf r,t)\,,
\end{split}
\ee
where $v_{ext}$ is the external potential including the applied synthesized laser field (in the velocity gauge) and electron-ion potential, $v_H$ is the Hartree part of the Coulomb electron-electron interaction,  $v_{xc}$ is exchange-correlation potential, $n(\mathbf{r},t)$ is the time dependent electron density defined as $n(\mathbf{r},t) = \sum_{i} |{\phi_i(\mathbf{r},t)}|^2$, where $\phi_i$ is the Kohn-Sham orbital associated with the index $i$ corresponding to both a band and a \textbf{k}-point index. The pseudopotential is not shown in Eq.~\ref{kseq} for simplicity.

We also consider the impact of nuclei dynamics during their interaction with the ultrashort pulse. For this purpose, the equation of motion for nuclei is described using Ehrenfest-TDDFT approach \cite{ehren}, where nuclei motions couple to the Kohn-Sham equations via a set of classical Newtonian equation,
$M_I \ddot{\mathbf R}_I = - \nabla_I [\sum_{i} \langle \phi_i | H_{KS} | \phi_i \rangle + E(\{R_I\})]$,
where $ H_{KS}$ and $\phi_i$ are, respectively, the Kohn-Sham Hamiltonian and electron wavefunction described in Eq.~\ref{kseq}, which are related to nuclei via electron-ion potential in $v_{ext}$; and $E(\{R_I\})$
is the classical energy of nuclei.
The initial velocities of the ions besides the initial system configuration  are taken from the CPMD simulations.

As shown earlier, our CPMD simulations capture the isotropic nature of the liquid system over 20~ps.
So, in order to correctly describe the interaction between the ultrashort laser pulse and the liquid,
we consider several different configurations of the CPMD time evolution,
which they denote as ``subsystems". The time interval between consecutive subsystems is on the order of several 10000 CPMD steps (a few ps) to ensure that the selected subsystems are independent.
The total electron current for each subsystem is computed from
\be
\label{ecurrent}
\mathbf{j}(t) = \sum_{i} \langle \phi_i | -i\mathbf{\nabla} + \frac{\mathbf{A}(t)}{c} | \phi_i \rangle\,,
\ee
where $\mathbf{A}(t)$ is the vector potential of the laser pulse and $c$ is the light speed.
In the next step, the high-harmonic spectrum is obtained by the Fourier transform of the electron current per water molecule,
averaged over different subsystems exposed to the same incident pulse
\be
\label{hhgeq}
HHG(\omega)= \sum_{k=x,y,z} \Big|FT \left(  \frac{\partial  \bar{j}_k(t)}{\partial t}  \right) \Big|^2\,,
\ee
where $k$ shows the mutually orthogonal directions in three-dimensional space, FT denotes a Fourier transform, $\bar{j}$ is average of electron current per water molecule, and $\omega$ is photon frequency.
Time-frequency analysis of high-harmonics is performed using the Gabor transform \cite{gabor}
\be
\label{gab}
\begin{split}
	G_{HHG}(\tau,\omega)= \Big| \int_{-\infty}^{\infty} dt \exp(\frac{-(t-\tau)^2}{2w^2})  e^{-i\omega t}
	\frac{\partial  \mathbf{\bar{j}}(t)}{\partial t}   \Big|^2\,,
\end{split}
\ee
where our time window is taken to be $w = 0.25$~fs for the laser wavelength of 1500~nm.

The exchange-correlation term in TDDFT calculations is described by the same GGA-revPBE as used in CPMD calculations. But, vdW interactions are not included, as they lead to no difference in the HHG spectra.
We consider a grid spacing of 0.3 bohr thorough TDDFT calculations, and a dense k-point grid of $5\times5\times5$. Also, we use aetrs algorithm to approximate the evolution operator and the time step of 0.2~a.u. within our TDDFT calculations.

Note that in this work we just considered the electron contributions to HHG,  which is computed from the time-dependent electronic current (see  Eq.~\ref{hhgeq}). We verified that the effects of ionic current are negligible.

\acknowledgments
The authors are grateful to the experimental group of Prof. H. J. W{\"o}rner  at ETHZ for useful discussions.
{\bf Funding:} This work is supported by the European Research Council (ERC-2015-AdG694097), the Deutsche Forschungsgemeinschaft (DFG) through the priority program QUTIF (SOLSTICE-281310551) and the Cluster of Excellence `CUI: Advanced Imaging of Matter'- EXC 2056 - project ID 390715994, Grupos Consolidados (IT1249-19), and 
the Max Planck - New York City Center for Non-Equilibrium Quantum Phenomena. The Flatiron Institute is a division of the Simons Foundation.
{\bf Author contributions:} Z.N. performed all the calculations. All authors discussed the results and contributed to the final manuscript.
{\bf Competing interests:} The Authors declare no Competing Interests.
{\bf Data and materials availability:}  All data needed to evaluate the conclusions in the paper are present in the paper and/or the Supplementary Materials.
\\

\newpage
\onecolumngrid

\title{Supplementary Information: An ab initio supercell approach for high-harmonic generation in liquids}

\author{Zahra Nourbakhsh}
\affiliation{Max Planck Institute for the Structure and Dynamics of Matter, Luruper Chaussee 149, 22761 Hamburg, Germany.}

\author{Ofer Neufeld}
\affiliation{Max Planck Institute for the Structure and Dynamics of Matter, Luruper Chaussee 149, 22761 Hamburg, Germany.}

\author{Nicolas Tancogne-Dejean}
\affiliation{Max Planck Institute for the Structure and Dynamics of Matter, Luruper Chaussee 149, 22761 Hamburg, Germany.}

\author{Angel Rubio}
\affiliation{Max Planck Institute for the Structure and Dynamics of Matter, Luruper Chaussee 149, 22761 Hamburg, Germany.}
\affiliation{The Hamburg Centre for Ultrafast Imaging, Hamburg, Germany.}
\affiliation{Nano-Bio Spectroscopy Group and ETSF, Departamento de Fisica de Materiales, Universidad del Pa\'is Vasco UPV/EHU, 20018, San Sebasti\'an, Spain.}
\affiliation{Center for Computational Quantum Physics (CCQ), The Flatiron Institute, 162 Fifth Avenue, New York, New York 10010, USA.}

\maketitle

\onecolumngrid

\setcounter{figure}{0}

\makeatletter 
\renewcommand{\thefigure}{S\@arabic\c@figure}
\makeatother


{\bf Supplementary Note 1: Radial distribution function between oxygen atoms in liquid water }  \\

\begin{figure}[h]
	\centering
	\includegraphics[width=0.5\linewidth]{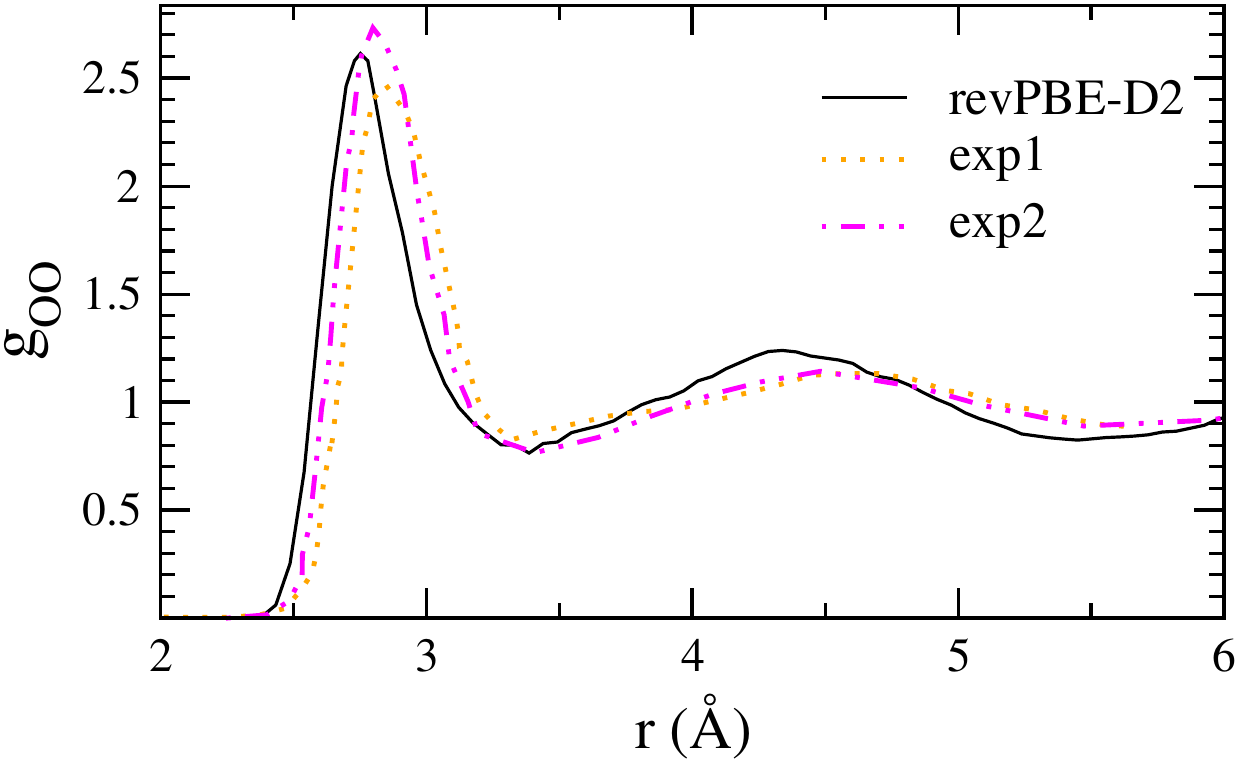}
	\caption{\label{rdf} Comparison of our calculated oxygen-oxygen radial distribution function (g$_\mathrm{OO}$) in liquid water with experiments; exp1 and exp2 are, respectively, obtained from neutron scattering [48] and joint x-ray/neutron diffraction [49]. }
\end{figure}

The usual way to estimate the accuracy of liquid water MD simulation is comparing the calculated radial distribution function (RDF) between oxygen atoms with   experiments [31]. 
Figure~\ref{rdf} displays O-O RDF obtained from our simulation which is calculated, after the equilibrium achieved, as time average over about 20~ps MD simulation.  This figure shows that our RDF is in a good consistent with the experimental measurements.

{\bf Supplementary Note 2: Toward the isotropic limit }  \\

\begin{figure}[h!]
	\centering
	\includegraphics[width=0.5\linewidth]{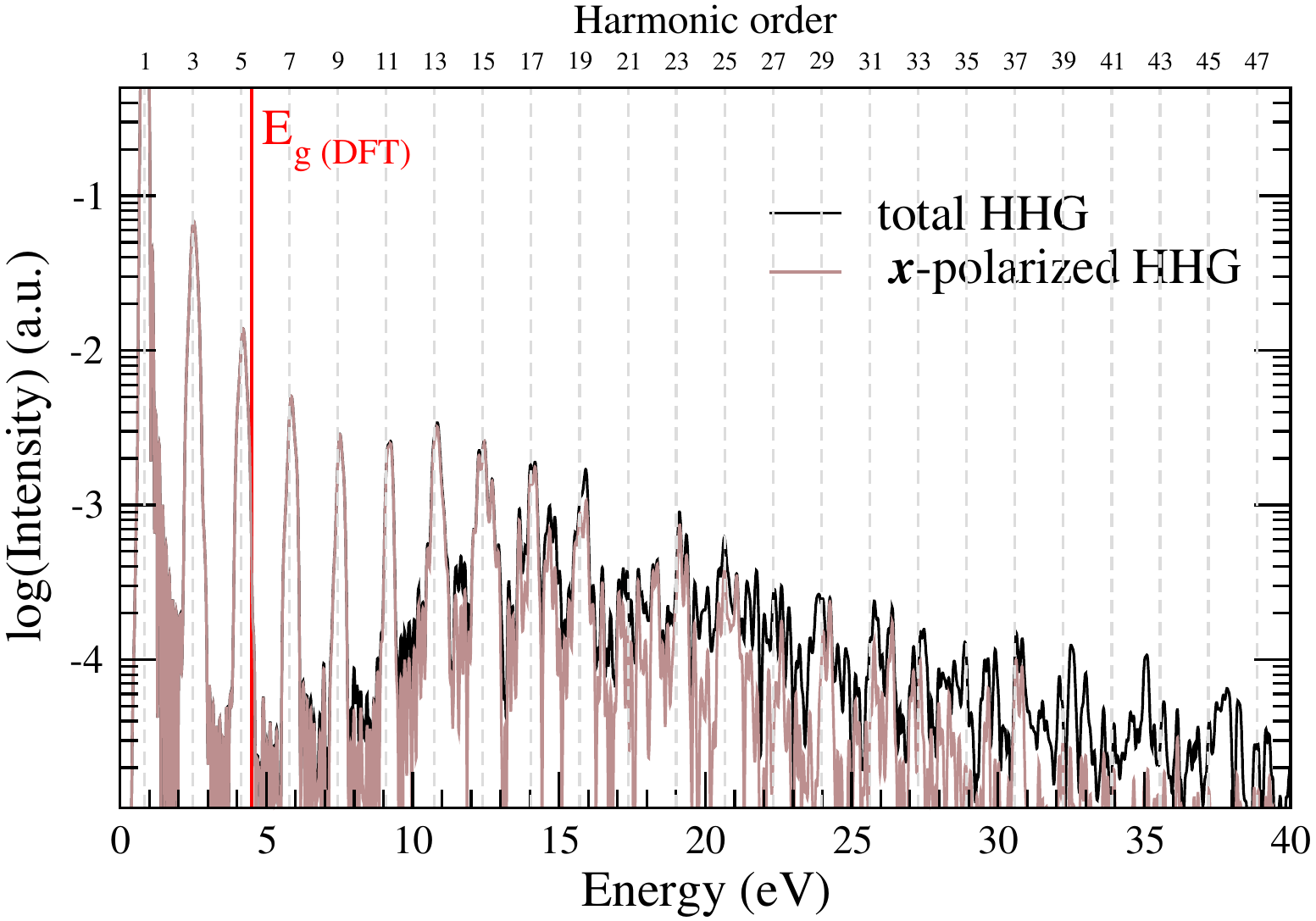}
	\caption{\label{totx} Comparison between the total HHG and harmonic emission along the incident laser pulse in the $x$ direction. By including more subsystems to fulfill the isotropic limit, the harmonic emission  in directions perpendicular to the pulse are suppressed.
	 }
\end{figure}

Similar to Fig.~3 in the main text, Fig.~\ref{totx} gives a comparison between total HHG and $x$-polarized HHG in liquid water. In this figure, the calculated results are averaged over 85 subsystems (approximately 5400 water molecule). The suppression of even harmonics as well as perpendicular emission at this spectrum marks the convergence of harmonic emission up to 15~eV. \\

{\bf Supplementary Note 3: Ice cutoff energy }  \\

Regarding the discrete band structure in solids, the cutoff energy definition for a HHG spectrum corresponding to a solid target is not so clean and one can define several cutoff energies. However, if We consider the cutoff energy as the spectral position where the intensity decreases by an order of magnitude at a given harmonic and beyond, averaged over a large bandwidth of several eV, the cutoff energy of crystalline structures in Fig.~5 will be between 15-20~eV which is higher than the cutoff of liquid.

\begin{figure}[h]
	\centering
	\includegraphics[width=0.5\linewidth]{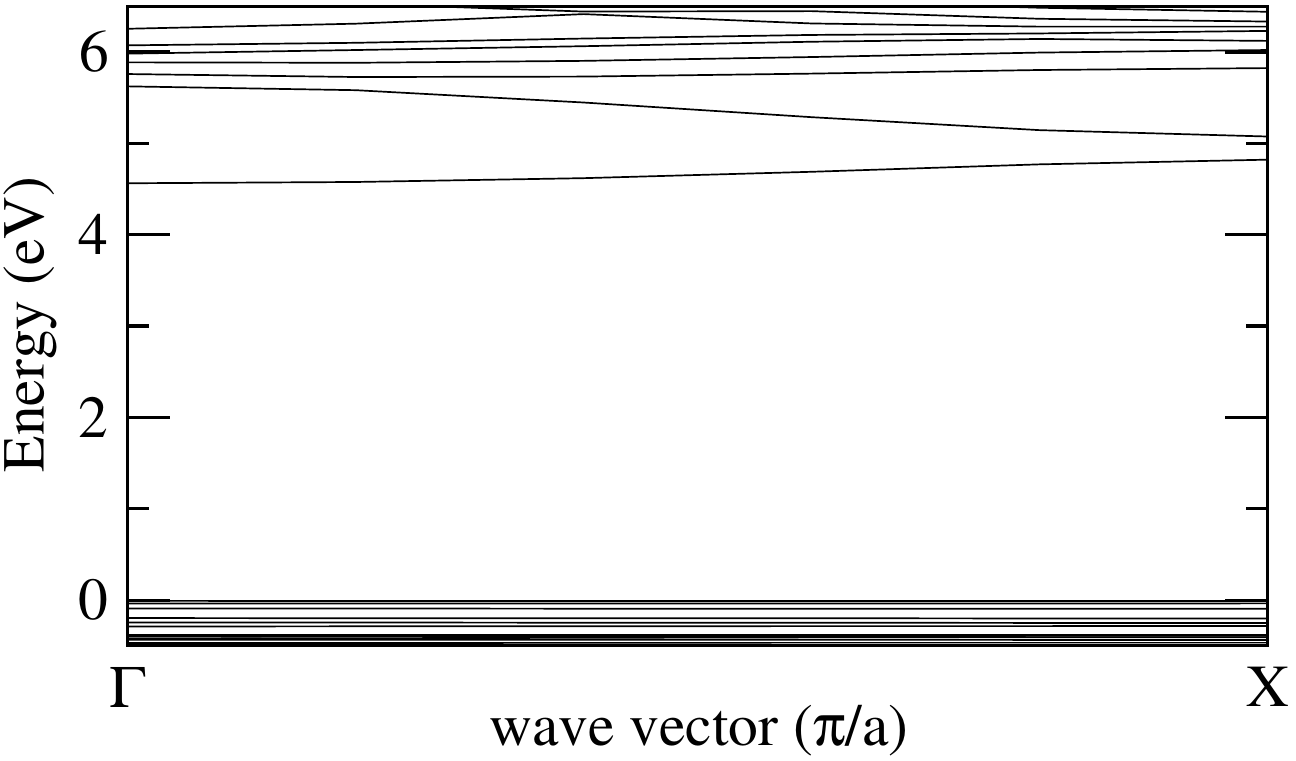}
	\caption{\label{band} The calculated electronic band structure for a typical subsystem of liquid water including 64 H$_2$O molecules. The top of the valence band is set to be zero. The occupied states are relatively flat while the hydrogen bonds between neighboring water molecules cause band dispersion in the conduction bands. }
\end{figure}

\begin{figure}
	\centering
	 \includegraphics[width=0.5\linewidth]{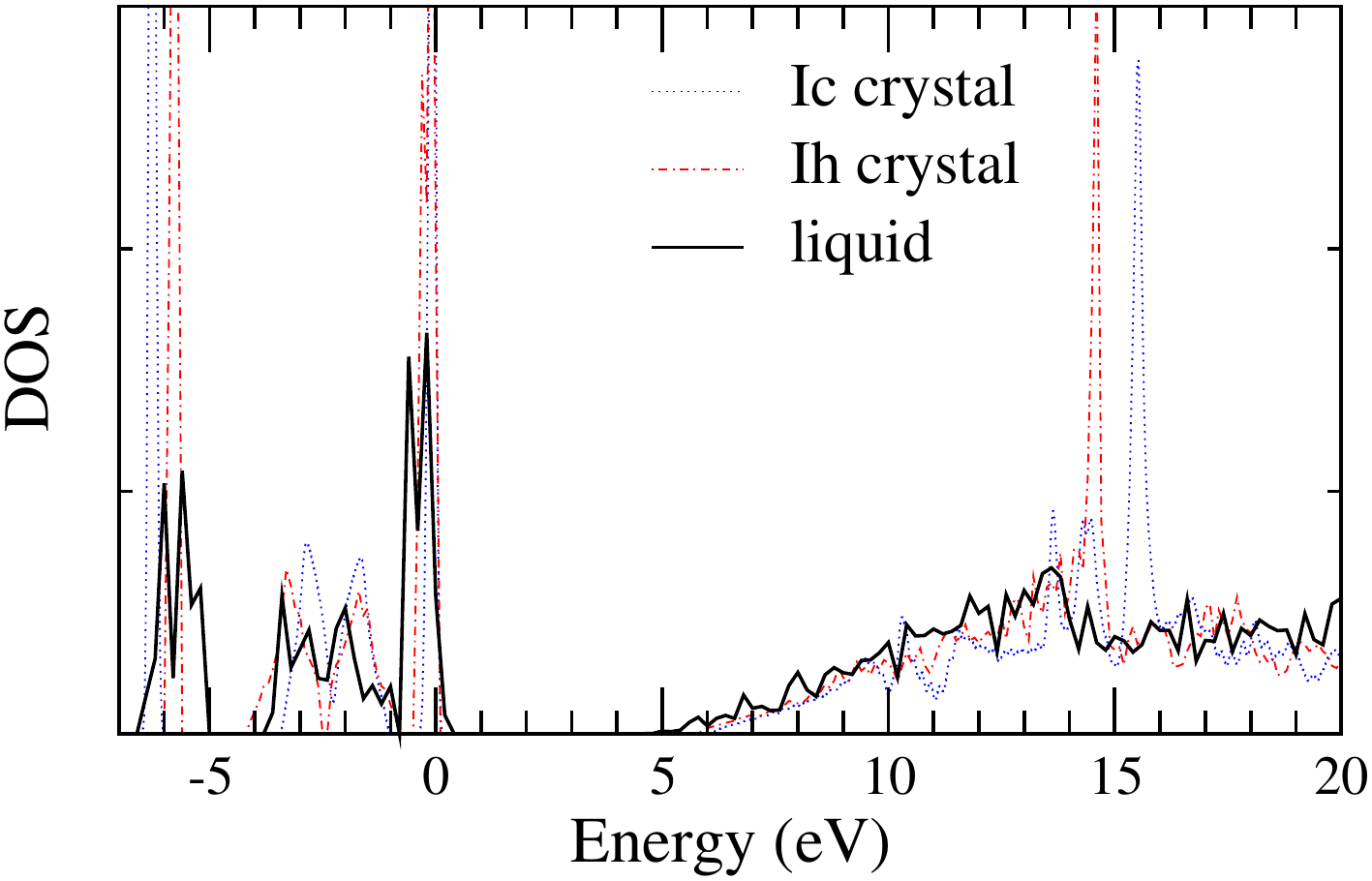}
	\caption{\label{dos} Comparison of density of states per H$_2$O molecule in liquid and crystalline phases of water. Top of the valence states are set to be zero.}
\end{figure}

\begin{figure*} 
	\centering
	\includegraphics[width=0.8\linewidth]{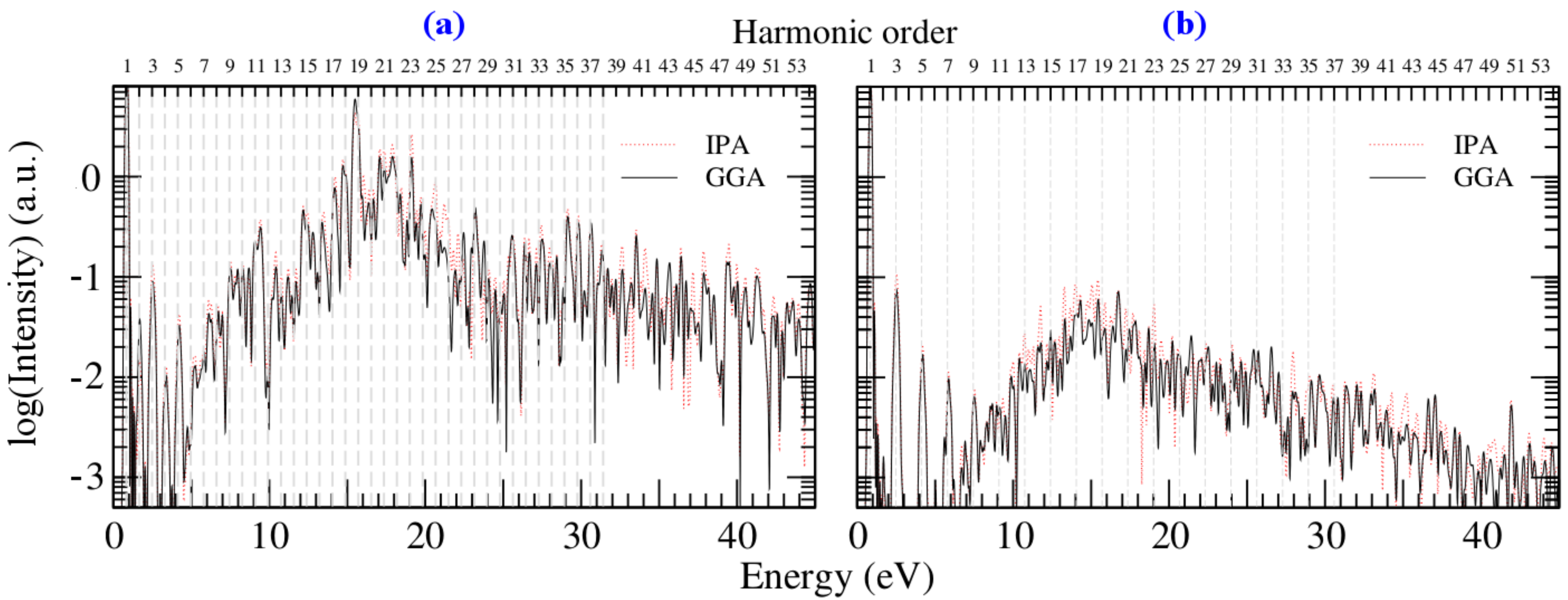}
	\caption{\label{locfld} Impact of electron-electron interactions on HHG from water; the considered interactions between electrons are in GGA-revPBE level. Comparisons are done for 
	(a) Ic crystal structure of water (introduced in the main text) and (b) a subsystem of liquid phase. 
The incident pulse is a linear polarized pulse as defined in the main text. 
Note that (b) gives a comparison just for one subsystem and the spectral is converged just for the low harmonics (see Fig.~\ref{nimp} in the main text). }
\end{figure*}

\begin{figure}
	\centering
	\includegraphics[width=0.5\linewidth]{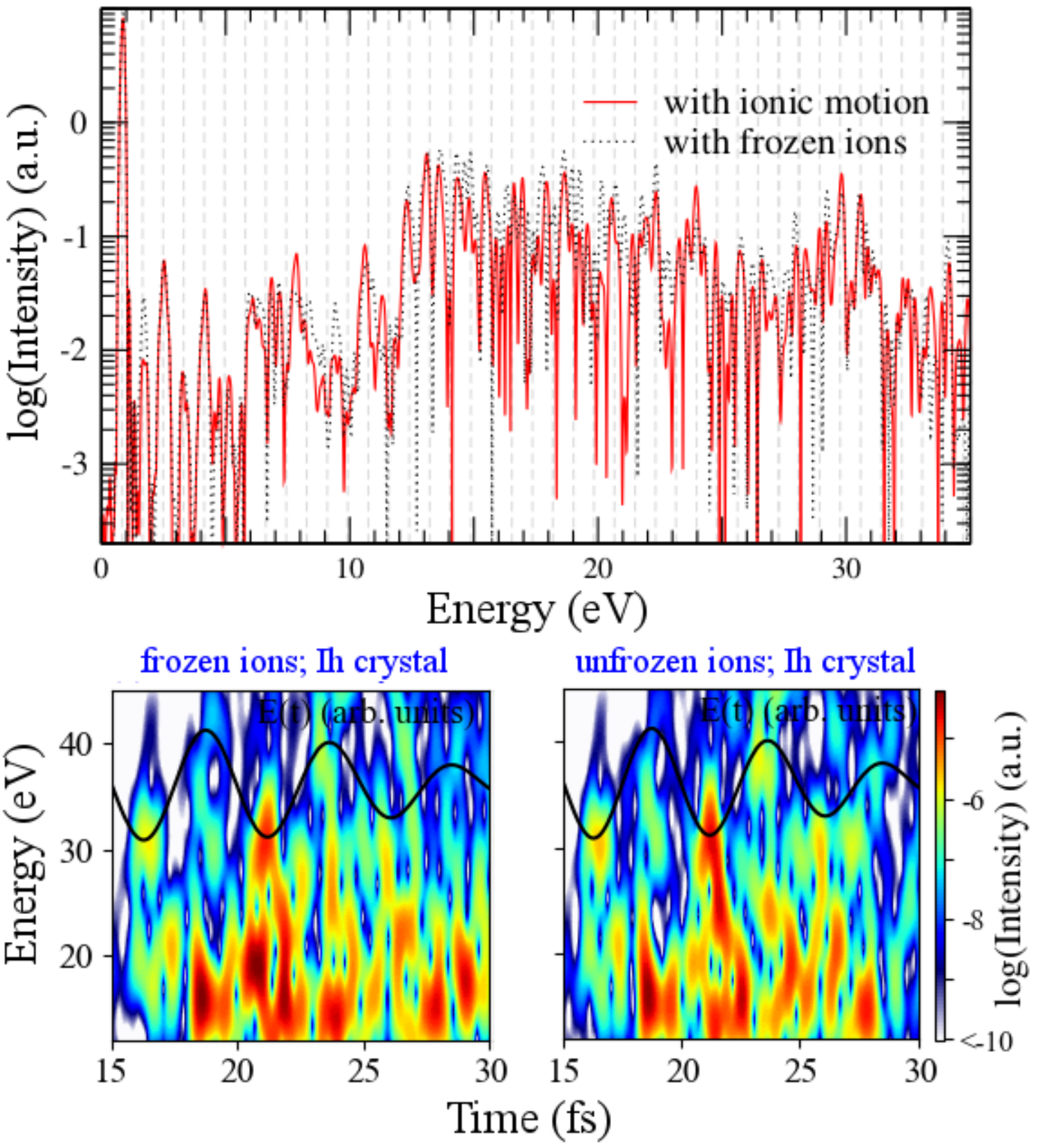}
	\caption{\label{ionmov2} Impact of ionic motions on HHG from Ih crystalline; the considered initial velocities are zero. The light polarization is along $z$ direction. The related time-frequency analysis are shown in the bottom panel; the sharp chirp in the right panel (with allowed ion dynamics) leads to extract a bright isolated attosecond pulse at the peak electric field. The time-dependent electric field is also displayed in the bottom panel.
The comparision between time-frequency analysis of ice and liquid water, displayed in the main text, reveals the different microscopic electron dynamics in these systems. While electron recombination in the crystalline structure of ice could be explained by inter- and intraband mechanisms, the time-frequency of liquid water shown in Fig.~6 indicates the scattering of excited electrons by other molecules. }
\end{figure}

\begin{figure*}
	\centering
	\includegraphics[width=0.8\linewidth]{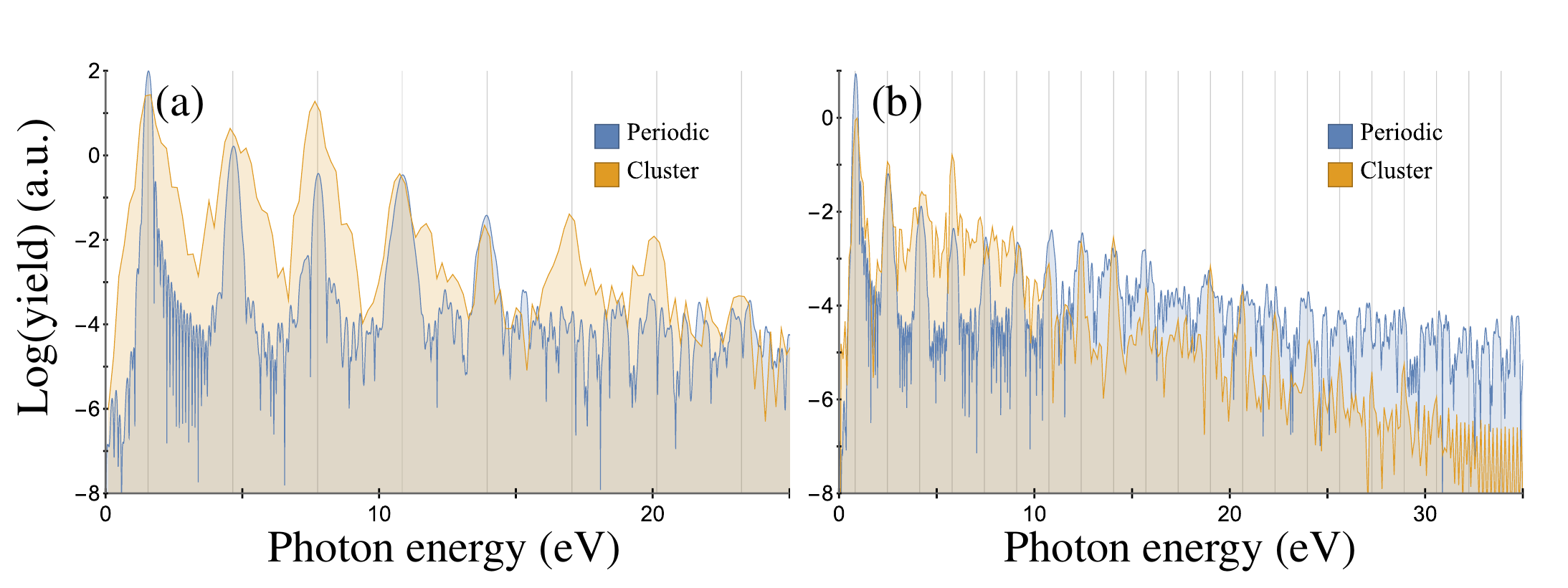}
	\caption{\label{clushhg} Liquid water HHG calculated using the cluster approach, compared to supercell results for two different incident pulse wavelengths of 800 and 1500~nm, in the same intensity of 20 TW/cm$^2$. Despite these quantitative differences, the cluster model qualitatively works well  and predict the cutoff energy as well as its independence from the wavelength in agreement with the supercell approach and experimental results, see also Fig.~3 in Ref.~39.  }
\end{figure*}

\end{document}